\documentclass[journal]{new-aiaa} 
\usepackage[utf8]{inputenc}

\usepackage{physics}
\usepackage{graphicx}
\usepackage{amsmath}
\usepackage[version=4]{mhchem}
\usepackage{siunitx}
\usepackage{longtable,tabularx}
\usepackage{todonotes}
\newcommand\extrafootertext[1]{%
  \bgroup
  \renewcommand\thefootnote{\fnsymbol{footnote}}%
  \renewcommand\thempfootnote{\fnsymbol{mpfootnote}}%
  \footnotetext[0]{#1}%
  \egroup
}

\title{Development and control of Görtler vortices in \\ the SD-2 serpentine diffuser \extrafootertext{Presented as Paper 2026-4623 at the 2026 AIAA Aviation Forum, San Diego, CA, June 8–12, 2026}}

\author{Akhil Nekkanti \footnote{Postdoctoral Fellow, AIAA Member}}
\affil{Center for Turbulence Research, Stanford University, Stanford, CA-94305, USA}
\author{Rahul Agrawal \footnote{Lead Software Engineer, Cadence Design Systems, AIAA Member.}} 
\affil{Cadence Design Systems, San Jose, CA-95134, USA \\ Center for Turbulence Research, Stanford University, Stanford, CA-94305, USA}

\author{Sanjeeb Bose \footnote{Distinguished Engineer, Cadence Design Systems and Adjunct Professor, ICME, Stanford
University. Associate Fellow AIAA.}}
\affil{Cadence Design Systems, San Jose, CA-95134, USA \\
Institute for Computational and Mathematical Sciences, Stanford University, Stanford, CA-94305, USA \\
Center for Turbulence Research, Stanford University, Stanford, CA-94305, USA}
\author{Parviz Moin \footnote{Franklin P. and Caroline M. Johnson Professor, Dept. of Mechanical Engineering, Stanford University. Fellow AIAA.}}
\affil{Center for Turbulence Research, Stanford University, Stanford, CA-94305, USA}

\begin{document}

\maketitle

\begin{abstract}
This study investigates and controls the flow through a serpentine diffuser using wall-modeled large-eddy simulations (WMLES), with emphasis on the formation of streamwise vortices, their contribution to flow distortion, and their suppression using opposition control. Following Agrawal et al., AIAA Journal (2026), simulations are performed at three Mach numbers, $Ma_{AIP} =0.36, 0.46,0.54$. Adverse pressure gradients and centrifugal forces induced by geometric features of the diffuser promote flow separation and lead to the formation of counter-rotating Görtler vortices. Downstream of the second bend, the transition from an elliptical to a circular cross-section drives these vortices toward the centerline along the spanwise direction, ejecting low-momentum fluid from the near-wall region into the core. A steady opposition-control strategy using wall-normal velocity forcing is employed to attenuate the streamwise vortices. The control strategy increases pressure recovery, reduces flow distortion, and substantially suppresses total-pressure fluctuations across all three Mach numbers. The maximum flow distortion is reduced by $\approx 50\%$ at the two lower Mach numbers and by $34\%$ at $Ma_{AIP}=0.54$. The variance of the total-pressure fluctuations is reduced by approximately two orders of magnitude near the center of the AIP and by a factor of two near the periphery. An input-output power analysis further shows a net positive power gain at all three Mach numbers, with a maximum control efficiency of approximately $65\%$ at the lowest Mach number, although the power efficiency decreases significantly at higher Mach numbers due to the larger actuation amplitudes required.
\end{abstract}

\section{Nomenclature}

{\renewcommand\arraystretch{1.0}
\noindent\begin{longtable*}{@{}l @{\quad=\quad} l@{}}
$\rho$ & density \\
$T$ & temperature \\
$p$ & pressure \\
$u_x,u_y,u_z$ & streamwise, spanwise, and vertical velocity, respectively \\
$\omega_x$ & streamwise vorticity \\
$c_{\infty}$ & speed of sound \\
$m$ & azimuthal wavenumber \\
$Ma_{AIP} $ &  Mach number at the AIP \\
$Ma_{t} $ &  Mach number at the throat \\
$L_x$ & streamwise length of the test section between the throat and the AIP \\ 
$P_{t_0}$ & reference inlet total pressure  \\
$P_{t_i}$ & total pressure  at the $i$-th probe\\
$\mathcal{P}$ & Power \\
$\langle \cdot \rangle $ & time-averaging operator \\
$\mathrm{AIP}$  & Aerodynamic Interface Plane \\
$\mathrm{D}_{AIP}$ & diameter of Aerodynamic Interface Plane \\ 
$\mathrm{A}_{AIP}$ & Area of Aerodynamic Interface Plane \\ 
$\mathrm{dPcP}_i$ & ring-averaged azimuthal distortion at the AIP \\
$\mathrm{dPcP}$ & face-averaged azimuthal distortion at the AIP \\
$\mathrm{PR}$ & temporally and face-averaged pressure recovery at the AIP \\
OC & Opposition control \\
WMLES & Wall-modeled Large-eddy simulation \\
LES & Large-eddy simulation \\
RANS & Reynolds-averaged Navier-Stokes \\
\end{longtable*}}

\section{Introduction}

Blended wing body configurations can aid the design of future military aircraft by integrating engines within the fuselage to improve aerodynamic efficiency and achieve a compact airframe with a high thrust-to-weight ratio. These configurations require compact inlets, including serpentine ducts \citep{winkler2017summary,brear2004flow}, where high-curvature bends generate large-scale vortices, shock waves, and flow separation \citep{ng2008relation}. Such unsteady flow features can reduce stall margin, and, in severe cases, lead to engine failure. In particular, non-uniform inflow conditions at the inlet exit can degrade aerodynamic performance, decrease engine operability, and increase the stress on compressor blades \citep{triantafyllou2015total}. \\

Flow through curved ducts, $90^{\circ}$ bend pipes, and S-shaped diffusers experiences radial pressure gradients and centrifugal forces that induce secondary flows along the walls and may lead to flow separation. This flow separation often increases the pressure losses, distorting the flow azimuthally at the aerodynamic interface plane (AIP)  \citep{wenzel1977analysis}. Counter-rotating vortices are formed inside the duct downstream of the bends. These vortices, reminiscent of Görtler vortices (also referred to as Dean vortices \citep{dean1927note,dean1928stream}) entrain the boundary layer fluid into the core, degrade pressure uniformity, and reduce pressure recovery \citep{brear2004flow,ng2008relation}.  It has been observed that such pressure distortions are influenced by the Mach number \citep{huang2024large,stratton2025development}; with 5.5\% higher pressure-recovery losses as $Ma_{AIP}$ increases from 0.3 to 0.55. In addition to mean pressure distortions, these inherently unsteady flows also exhibit peak dynamic distortions \citep{zachos2016flow,macmanus2017complex}, which can trigger compressor instabilities even when mean levels remain within acceptable limits. The dynamical behavior of these flow phenomena is not fully understood since it is difficult to analyze three-dimensional flow structure upstream of the AIP non-intrusively in experiments.  \\

Numerous experiments \citep{brear2004flow,ng2008relation,harrison2013active} have studied the mean flow distortions through serpentine diffusers. However, only a few numerical studies have been conducted to physically characterize the mechanism of flow distortion. Given the high Reynolds numbers of these flows, direct numerical simulations are also generally intractable. \citet{lakebrink2018numerical} have shown that Reynolds-averaged Navier-Stokes (RANS) models overpredict mean flow recovery, and while detached eddy simulations (DES) provide relatively more accurate results, their predictions deteriorate with increasing Mach number. Similarly, \citet{lakebrink2019toward} and \citet{stahl2023modal} performed coarsely-gridded wall-modeled large eddy simulations (WMLES) of the SD2 diffuser \citep{burrows2019control,burrows2021active} and reported overpredictions of pressure recovery at the exit plane. However, recently, \citet{agrawal2025predictions} showed that WMLES can predict the average recovered pressure and flow distortion metrics within the experimental error bounds when the grids are refined to capture the dominant pressure-gradient imposed length scales \citep{agrawal2024reynolds}. The present study builds on these simulations and examines the spatial development of counter-rotating vortices inside the diffuser, their role in generating mean distortion, and the sensitivity of these vortices to Mach number. \\

Several previous studies have explored passive and active flow control strategies to reduce distortion and enhance pressure recovery in curved ducts. Among passive methods, vortex generators \citep{brown1968subsonic,ng2008relation} have been shown to delay boundary layer separation by inducing vortices that draw high-momentum fluid from the core into the boundary layer, countering secondary flows \cite{lin2002review}. Although passive techniques are robust, they cannot adapt to changing flow conditions in real time, limiting their overall effectiveness. In contrast, active flow control (AFC) enables real-time tuning of control parameters \cite{wojewodka2018review}. AFC techniques such as variable directional tangential blowing \cite{ball1985tests,ng2008relation,harrison2013active} and fluidic injection  \cite{burrows2019control, burrows2021active,burrows2021experimental, scribbenEffectivenessSerpentineInlet2004,gartner2019mitigation,burrows2025controlled} offer improved performances with lower drag penalties. The physical mechanism involves injection of high-momentum fluid and suction of low-momentum fluid, which enhances the flow’s resistance to adverse pressure gradients \citep{huang2024large}. For serpentine diffusers, \citet{gartner2019mitigation} demonstrated improved total-pressure recovery using pulsed jets, sweeping jets, and blowing slots, while \citet{burrows2019control,burrows2021experimental} achieved a 60\% reduction in AIP flow distortion by suppressing secondary-flow vortices with fluidic-oscillating jets. In a serpentine duct with a cowl inlet, fluidic injection provided a substantial increase in pressure recovery from 85\% to 93\% \citep{burrows2025controlled}. \citet{scribbenEffectivenessSerpentineInlet2004} employed air injection through microjets at 1\% of the mass flow rate to reduce inlet distortion at the AIP by 70\%, while increasing total pressure recovery by 2\%. While these strategies effectively reduce mean flow distortion, comparatively few studies have examined their effectiveness in reducing maximum distortion levels. The present study addresses these gaps by quantifying the power efficiency of flow control through an input-output power analysis and by evaluating its ability to reduce both time-averaged and, in particular, maximum flow distortion. \\  

In this work, we utilize the opposition control (OC) methodology, proposed by \citet{choi1994active}; OC is an active flow control strategy that imposes velocities at the wall in opposition to those measured at a sensing plane located within the flow. Using this strategy, \citet{choi1994active} demonstrated up to 25\% reduction in skin-friction drag in a turbulent channel flow. The underlying principle of opposition control is to weaken streamwise vortices in the near-wall region by prescribing wall velocities that oppose the induced motion of these vortical structures. As expected, OC is sensitive to the sensing-plane location \cite{hammond1998observed,chung2011effectiveness}. \citet{hammond1998observed} showed that a sensing plane located too far from the wall can have adverse effects, while \citet{chung2011effectiveness} demonstrated that there exists a maximum amplitude beyond which opposition control becomes less effective. \citet{chang2002viscous} showed that the effectiveness of opposition control decreases at higher Reynolds numbers, with drag reduction decreasing from 25\% to 19\% as $Re_{\tau}$ increased from 100 to 1000. \citet{pamies2007response} performed targeted opposition control that suppresses only sweep events, resulting in 61\% drag reduction. To date, most studies have employed opposition control under zero-pressure-gradient conditions and have not investigated its effectiveness in complex geometries. Moreover, near the flow separation, the local friction Reynolds number may be small, making OC a viable approach to control flow distortion. Thus, herein, we aim to assess the efficacy of steady opposition control in suppressing the large Görtler vortices that lead to flow distortions in the SD-2 duct. We highlight that the current approach has a limited spatial extent of control and differs from previous studies by using time-invariant actuation (i.e., based on mean velocities) rather than instantaneous fluctuations. \\

The remainder of this article is organized as follows: Section \ref{methodology} provides a brief overview of the governing equations and the computational solver. Section \ref{results} presents the development of the streamwise vortices inside the diffuser. Section \ref{sec:control} presents the opposition-based flow-control methodology and evaluates its effectiveness in improving pressure recovery, reducing flow distortion, and quantifying the associated power efficiency. Section \ref{sec:M_effects} examines the effectiveness of the control strategy at higher Mach numbers. Finally, concluding remarks are offered in Section \ref{conclusions}.

\section{Governing equations and computational solver} \label{methodology}

This work utilizes the charLES flow solver to perform large eddy simulations\footnote{from Cadence Design Systems, San Jose, CA}: this solver is an explicit, unstructured, finite-volume code which solves the filtered compressible Navier-Stokes equations (see below), and is second-order accurate in space, and third-order accurate in time. For further details on the solver and relevant validation studies, the reader is referred to  \cite{bres2017unstructured,goc2023wind,agrawal2024reynolds,agrawal2025predictions}. The resulting system of equations is:  

\begin{align}
\frac{\partial \bar{\rho}}{\partial t} + \frac{\partial (\bar{\rho}\,\tilde{u}_j)}{\partial x_j} &= 0, \label{eq:continuity}\\
\frac{\partial (\bar{\rho}\,\tilde{u}_i)}{\partial t}
+ \frac{\partial (\bar{\rho}\tilde{u}_i\tilde{u}_j) }{\partial x_j}
&= - \frac{\partial \bar{p}}{\partial x_j} + \frac{\partial \tilde{\tau}_{ij}}{\partial x_j} -\frac{\partial \tilde{\tau}^{\mathrm{sgs}}_{ij}}{\partial x_j}, \label{eq:momemtum} \\
\frac{\partial \bar{E}}{\partial t}
+ \frac{\partial (\bar{E} \tilde{u}_j) }{\partial x_j} 
&=  - \frac{\partial \bar{p}\tilde{u}_j }{\partial x_j} + \frac{\partial \tilde{\tau}_{ij}\tilde{u}_i }{\partial x_j}  - \frac{\partial \tilde{\tau}^{\mathrm{sgs}}_{ij}\tilde{u}_i }{\partial x_j} - \frac{\partial \tilde{Q}^{\mathrm{sgs}}_{ij}}{\partial x_j}  + \frac{\partial}{\partial x_j}\bigg(\kappa\frac{\partial \bar{T}}{\partial x_j}\bigg),
\label{eq:energy}
\end{align}
where $\bar{(\cdot)}$ and $\tilde{(\cdot)}$ denote the filtered and Favre-averaged quantities, respectively. Here, $u_i$ represents the $i$-th component of velocity, $p$ the pressure, $\rho$ the density, $E$ the total energy, $T$ the temperature, and $\tau_{ij}$ and  $\tau_{ij}^{\mathrm{sgs}}$ are the viscous and subgrid stresses, respectively. Here, we invoke the dynamic Smagorinsky model\citep{moin-dynamic} to approximate the subgrid-scale stresses. The molecular Prandtl number is prescribed as a  constant, $Pr= 0.7$. The subgrid heat flux is computed using a constant turbulent Prandtl number approximation ($Pr_t = 0.9$). The equilibrium wall model \citep{cabot2000approximate} is employed to approximate the boundary conditions on the solid walls of the diffuser. Further details of the equilibrium wall-model formulation employed in this study are provided in \citet{lehmkuhl2018large} and \citet{goc2021large}.\\

Recently, \citet{agrawal2025predictions} performed WMLES of flow through the SD-2 duct at three Mach numbers $Ma_{\mathrm{AIP}}=0.36, 0.46$, and 0.54. In line with the \emph{a-priori} estimates \citep{agrawal2024reynolds}, it was demonstrated that a grid (termed 'L7' in \citet{agrawal2025predictions}) resolving the viscously imposed pressure-gradient length scale to $\Delta^+_p \sim 10$ was necessary to capture the pressure recovery and flow distortion at the AIP within experimental uncertainty bounds. Following this work, we study the same mass flow rates on the L7 grid (containing three billion control volumes). The flow is simulated for 40 convective time units, which is sufficiently long to flush out initial flow transients, and collect statistics, including estimates of peak azimuthal distortion (see \citet{agrawal2025predictions} for dominant frequencies of changes in peak distortion values in the SD-2 diffuser). \\

\begin{figure}
            {\includegraphics[trim={0.0cm 4.25cm 0cm 5.8cm},clip,width=1.0\textwidth]{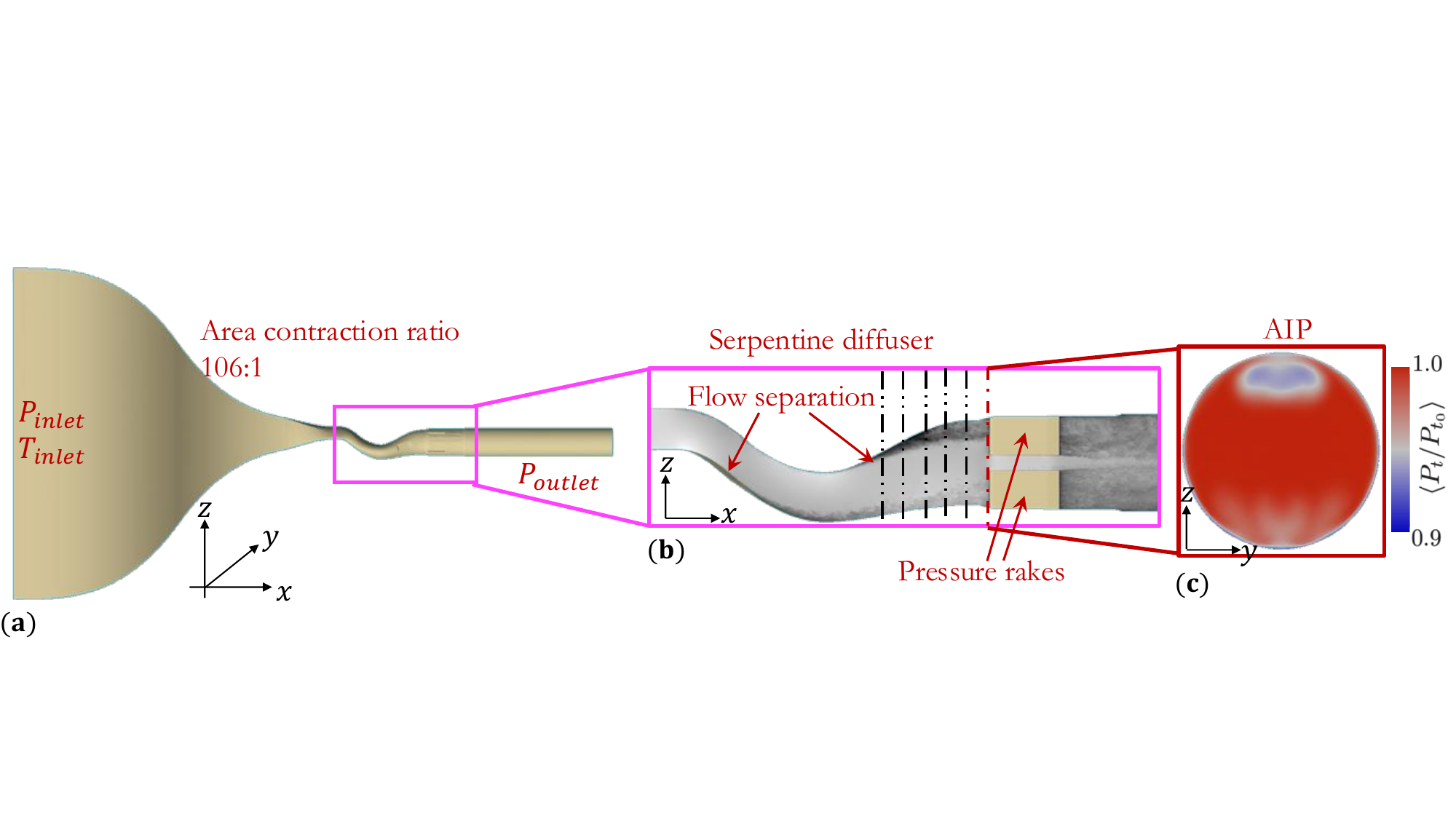}}
      	\caption{ (a) Schematic of the computational setup, which includes  a contracting inlet of area ratio 106:1, leading to the test section of the serpentine diffuser followed by a cylindrical section. (b) Streamwise velocity shown on $x-z$ plane. (c) Total pressure at the AIP.  Black dash-dotted lines denote the streamwise planes that will be shown in Figs. \ref{Fig2} and \ref{Fig3}.} \label{Fig1}
\end{figure}

The computational setup is shown in Fig.~\ref{Fig1}(a), herein, the domain consists of a bellmouth contraction inlet with an area ratio of 106:1 that accelerates the flow into the diffuser (test section). Downstream of the diffuser is a cylindrical section that houses the pressure probe rakes and acts as the flow outlet. The coordinates $x$, $y$, and $z$ denote the streamwise, spanwise, and vertical directions, respectively. Fig.~\ref{Fig1}(b) presents a side-view of the instantaneous streamwise velocity contours, highlighting the flow separation at the bends of the diffuser, at $x/D_{AIP}\approx -3.00$ and $x/D_{AIP} \approx-1.60$, respectively. The adverse pressure gradients combined with the curvature at these bends lead to separation, which consequently causes azimuthal flow distortion at the AIP (see Fig.~\ref{Fig1}(c)). The flow separation induced by the second bend along the upper wall exerts a stronger influence, likely due to its spatial proximity, on the distortion at the AIP, manifested by the blue contours in panel (c). Flow separation along the bottom wall also causes distortion and a loss of total pressure, albeit to a lesser extent.  

\section{Characterization of streamwise vortices and flow distortion} \label{results}

This section describes the flow features observed inside the SD-2 diffuser and their dependence on Mach number. The streamwise vorticity and velocity are examined at multiple streamwise stations. The spatial development of these flow structures and their relationship to regions of low total pressure are  characterized. Furthermore, to relate the pressure fluctuations to the Reynolds-stress events, a quadrant analysis \cite{wallace1972wall} of the velocity fluctuations at the AIP is performed and presented in Appendix I.

\subsection{Streamwise flow features}

\begin{figure}
            {\includegraphics[trim={0.0cm 4.75cm 0cm 2.7cm},clip,width=1.0\textwidth]{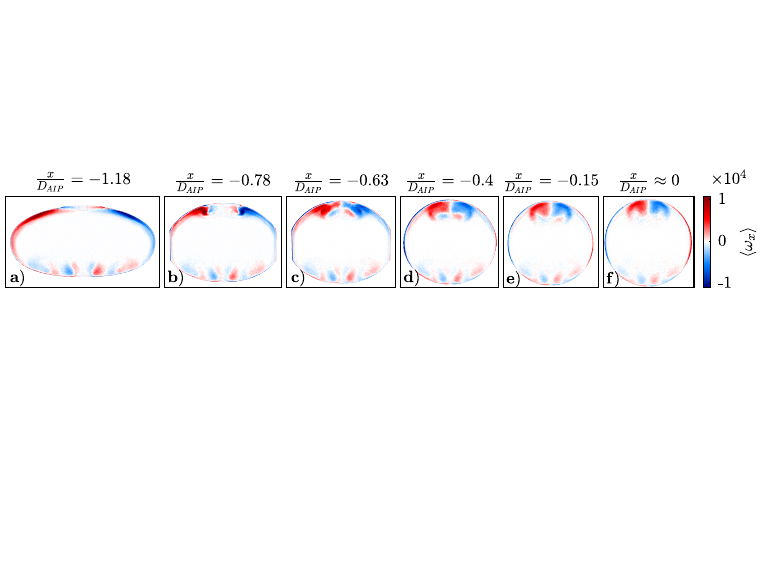}}
      	\caption{Mean streamwise vorticity at different axial planes for the flow inside the SD-2 diffuser at $Ma_{AIP} \approx 0.36$ conditions. The locations are (a) $x/D_{AIP}=-1.18$; (b) $x/D_{AIP}=-0.78$; (c) $x/D_{AIP}=-0.63$; ($d$) $x/D_{AIP}=-0.4$; ($e$) $x/D_{AIP}=-0.15$; ($f$) $x/D_{AIP} \approx 0$ (AIP). These streamwise planes are marked by dash-dotted lines in the full computational setup shown in Fig. \ref{Fig1}b. 
        \label{Fig2}}
\end{figure}

\begin{figure}
            {\includegraphics[trim={0.0cm 4.75cm 0cm 2.7cm},clip,width=1.0\textwidth]{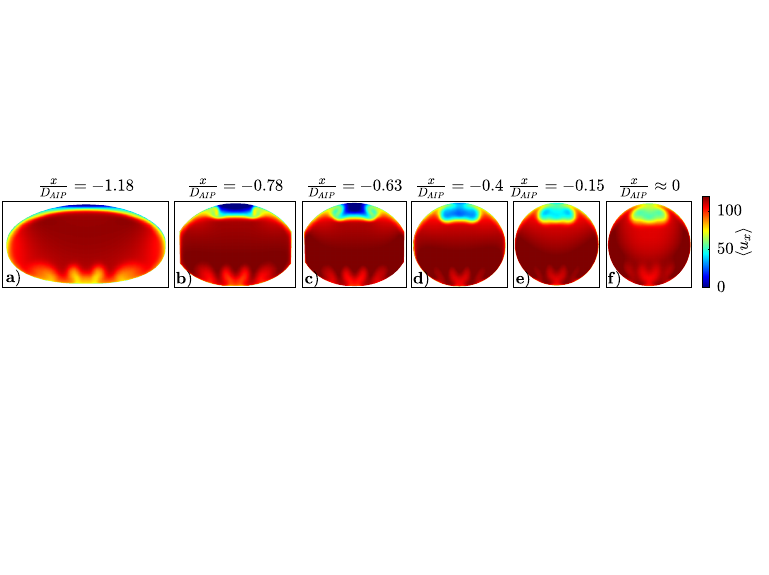}}
      	\caption{Mean streamwise velocity at different axial planes for the flow inside the SD-2 diffuser at $Ma_{AIP} \approx 0.36$ conditions. The locations are (a) $x/D_{AIP}=-1.18$; (b) $x/D_{AIP}=-0.78$; (c) $x/D_{AIP}=-0.63$; ($d$) $x/D_{AIP}=-0.4$; ($e$) $x/D_{AIP}=-0.15$; ($f$) $x/D_{AIP} \approx 0$ (AIP). \label{Fig3} }
\end{figure}

First, we consider the lowest Mach number, $Ma_{AIP} =0.36$; and the effects of Mach number are discussed later. The streamwise vorticity, and the velocity at the AIP are presented in Figs.~\ref{Fig2} and \ref{Fig3} respectively. Fig.~\ref{Fig2} presents the mean streamwise vorticity at several axial cross-sections spanning $-1.18 \leq x/D_{AIP} \leq 0.0 $, spanning the separated region close to the second bend leading up to the AIP. The vorticity contours demonstrate the presence of counter-rotating vortices along the upper wall, which develop as the boundary layer separates. These Görtler vortices arise from radial pressure gradients induced by wall curvature. As the flow evolves downstream, the counter-rotating vortices move toward the centerline ($y/D_{AIP}=0$) and into the core region. The vortices appear to be quasi-developed by $x/D_{AIP}=-0.15$, beyond which they become spatially fully developed and exhibit little variation in their structure along the streamwise direction. Additionally, pairs of counter-rotating vortices of lower amplitude are observed along the lower wall, likely originating from the boundary-layer separation induced by the first bend. Note that the sharp corners visible in some cross-sections, for example in Fig.~\ref{Fig2}(b,c), arise from the optical PIV windows used in the experiments and further amplify the streamwise vortices.

Fig.~\ref{Fig3} presents the mean streamwise velocity at the same cross-sectional planes. At $x/D_{AIP}=-1.18$, a region of low streamwise velocity is observed along the top wall, which localizes closer to the centerline downstream. This low-momentum region, consistent with the detachment of the boundary layer, lies between the counter-rotating vortices, as evidenced in Figs.~\ref{Fig2}(b) and \ref{Fig3}(b). This region is subsequently advected from the near-wall into the core by the counter-rotating vortices. At the AIP, the low-momentum region coincides with the area of total pressure loss. A similar behavior is observed along the bottom wall. This process demonstrates that the Görtler vortices eject the low speed fluid from the wall into the core leading to distortion. A detailed quadrant analysis is presented in Appendix I, that decomposes the Reynolds shear stress into inward and outward motions, ejection and sweep. Conditional statistics further show that ejection events coincide with the largest total-pressure losses. \\  

\begin{figure}
\centering
            {\includegraphics[trim={0.0cm 4.7cm 0cm 2.8cm},clip,width=1.0\textwidth]{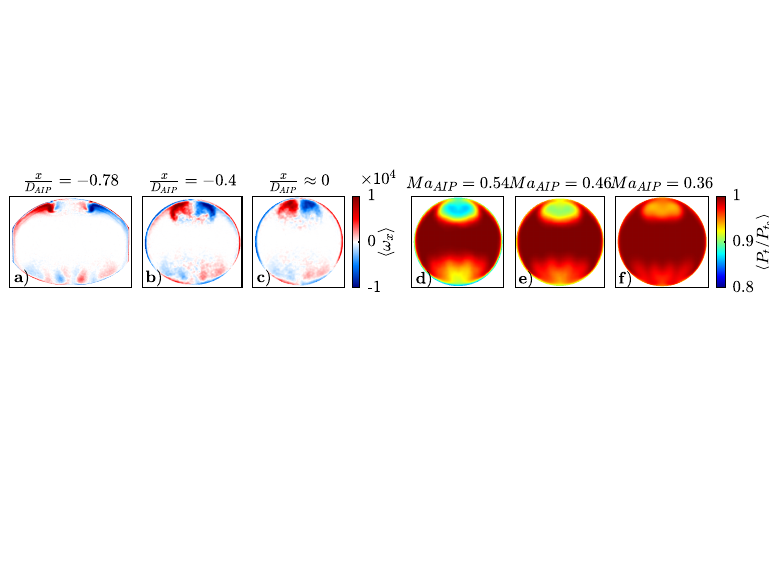}}
      	\caption{Mean streamwise vorticity at different axial planes for $Ma_{AIP}=0.54$ flow inside the SD-2 diffuser: (a) $x/D_{AIP}=-0.78$; (b) $x/D_{AIP}=-0.4$; (c) $x/D_{AIP}=0$ (AIP). Total pressure at the AIP for different Mach number conditions: ($d$) $Ma_{AIP}=0.54$; ($e$) $Ma_{AIP}=0.46$; ($f$) $Ma_{AIP}=0.36$. \label{Fig3b} }
\end{figure}

Next, we consider the $Ma_{AIP} = 0.54$ flow to examine the effect of the Mach number on the development of the streamwise vortices. Fig.~\ref{Fig3b}(a–c) presents the mean streamwise vorticity at three stations. Qualitatively, the spatial features of the streamwise vortices are similar to those observed at the lower Mach number; however, at the higher Mach number, the vorticity amplitudes are higher. Fig.~\ref{Fig3b}(d–f) presents the total pressure at the AIP across the Mach number range; it is apparent that at the higher Mach numbers, the pressure recovery is smaller, and the lowered-recovery regions are localized in the vicinity of the Görtler vortices. Qualitatively, the pressure recovery patterns are similar across the Mach numbers, which suggest that while the Mach number amplifies the distortion, the underlying vortex dynamics is similar.  \\


\section{Opposition control} \label{sec:control}

\begin{figure}[!t]
\centering
{\includegraphics[trim={0.0cm 6.4cm 0cm 1.3cm},clip,width=1.0\textwidth]{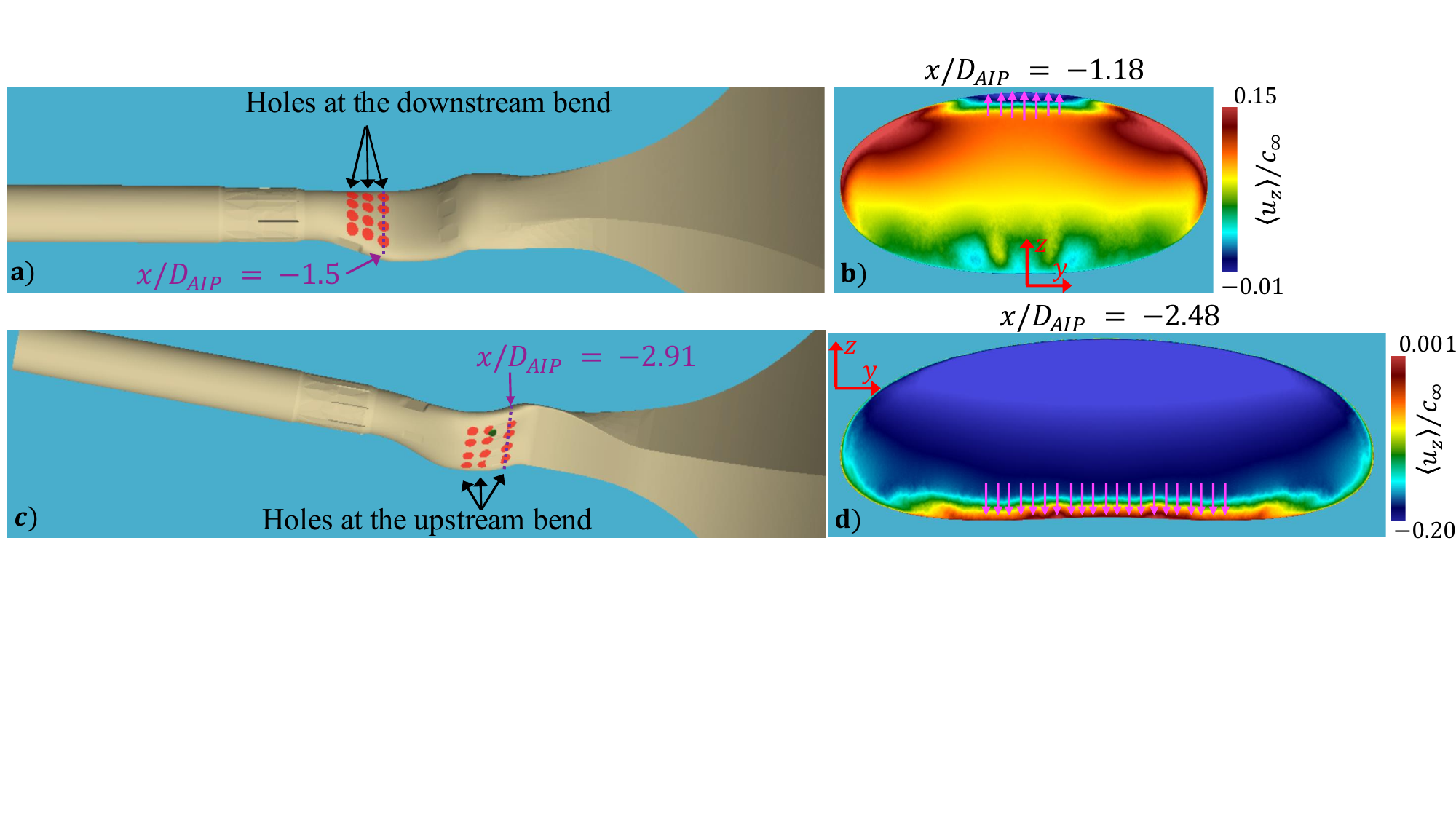}}
\caption{Schematic of opposition control: (a,c) holes at the downstream and upstream bends, respectively; (b,d) vertical-velocity contours at $x/D_{AIP}=-1.18$ and $-2.48$, respectively, for the uncontrolled $Ma_{AIP}=0.36$ flow. Holes have diameter $d_h/D_{AIP}=0.236$. Upstream rows are centered at $x/D_{AIP}=-2.91,-2.48,-2.13$, with four holes per row at $y/D_{AIP}=\pm0.20,\pm0.51$; downstream rows are at $x/D_{AIP}=-1.50,-1.18,-0.87$, with holes at $y/D_{AIP}=\pm0.20,\pm0.43$.}
        \label{Fig10}
\end{figure}

Next, based on the analyses presented thus far, an opposition control strategy is performed to suppress Görtler vortices and improve pressure recovery. First, a parametric study is performed in $Ma_{AIP}=0.36$ to examine the effects of control amplitude on mean total-pressure recovery, mean distortion and maximum distortion, as well as to assess the overall effectiveness of the control strategy. The results at this Mach number are discussed in detail in this section. The corresponding results for $Ma_{AIP}=0.46$ and $0.54$ are presented subsequently in section \ref{sec:M_effects}.

At each bend, twelve control holes, each with diameter $d_{h}/D_{AIP} \approx 0.24$, are arranged in three streamwise rows of four holes each, with opposition control applied as a boundary condition at these locations. The holes shown in Fig.~\ref{Fig10}(a)-(b) are placed to encompass the spanwise extent over which flow separation occurs. At the upstream bend, the three rows of holes are centered at $x/D_{AIP} = -2.91, -2.48,$ and $-2.13$ with spanwise locations $y/D_{AIP} = \pm0.51, \pm0.20$. At the downstream bend, the rows are centered at $x/D_{AIP} = -1.50, -1.18,$ and $-0.87$ with spanwise locations $y/D_{AIP} = \pm0.43, \pm0.20$. At each bend, the first row of holes (located at $x/D_{AIP} = -2.91$ and $-1.50$ respectively) is positioned at the onset of flow separation. Although not shown, the effectiveness of the control was found to be sensitive to the hole placement, in particular, the row closest to the AIP. Consequently, we do not pursue placing holes in the straight section of the diffuser ($-0.20 \lesssim x/D_{AIP} \lesssim 0.025$), where the flow is originally separated. The hole size relative to the local boundary-layer thickness varies across the control region, as the boundary layer is thinner near the onset of flow separation and thickens downstream. The boundary-layer thickness, $\delta_{99}$, is defined by $u(x,z=\delta_{99})=0.99U_e(x)$, where $U_e$ is the freestream velocity. The ratios of hole size to boundary-layer thickness, from the first to the third row, are $d_h/\delta_{99}=2.72,0.52,0.48$ at the first bend and $4.34,2.06,0.50$ at the second bend. \\

Fig.~\ref{Fig10} (c,d) shows the mean vertical velocity for the unperturbed $Ma_{AIP}=0.36$ flow, on streamwise planes $x/D_{AIP} = -1.18$ and $-2.48$, respectively.  Near the top wall at $x/D_{AIP} = -1.18$ and near the bottom wall at $x/D_{AIP} = -2.48$, the vertical velocity is \emph{opposite} to that in the core region. Additionally, a sharp spanwise gradient $\partial \langle u_z\rangle/\partial y$ is observed near the curved walls. This variation in vertical velocity contributes to the generation of streamwise vorticity through the term $\partial u_z/\partial y$ (where $z$ is the vertical coordinate, positive upward, and $y$ is the spanwise direction). Therefore, we impose steady-suction at both bends to remove the low-momentum fluid and attenuate the streamwise vortices. A schematic of the suction application is shown in Fig. \ref{Fig10}(c,d) by magenta arrows.  


\begin{figure}
\centering
{\includegraphics[trim={0.0cm 3cm 0cm 1.8cm},clip,width=1.0\textwidth]{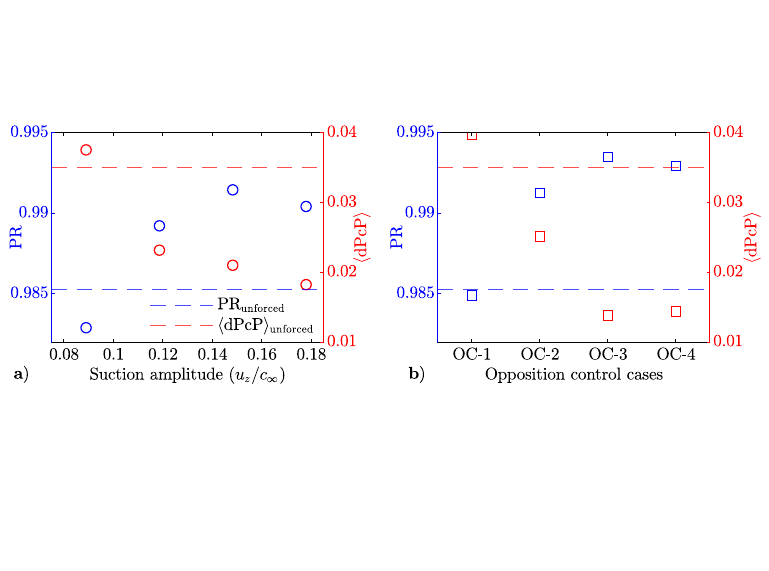}}
      	\caption{ Effect of ($\textbf{a}$) suction amplitude (vertical velocity normalized by the speed of sound) and ($\textbf{b}$) different opposition control cases, reported in Table \ref{table:OC-cases}, on pressure recovery and time-averaged flow distortion, $\langle \rm{dPcP} \rangle$. The unperturbed case is denoted by dashed lines.} \label{Fig11}
\end{figure}

\subsection{Impact of suction on pressure recovery and flow distortion}
First, we examine the effect of the uniform suction amplitude (applied at all holes) on the time-averaged, recovered face-averaged pressure (PR), face-averaged pressure distortion ($\langle \rm{dPcP}\rangle$), and maximum pressure distortion ($\rm{\max(dPcP})$)  at the AIP. The mathematical definitions of PR and $\langle \rm{dPcP}\rangle$ can be found in \cite{sae1420,agrawal2025predictions} and are reiterated here for completeness:
\begin{align}
    \mathrm{PR} & = \sum^{\mathrm{Number \, of \, rings}}_{i=1} \frac{\langle P_{t_i} \rangle}{P_{t_0}},  \quad \text{and} \\
    \mathrm{dPcP} & \equiv \sum^{\mathrm{Number \, of \, rings}}_{i=1} \mathrm{dPcP}_i = \sum^{\mathrm{Number \, of \, rings}}_{i=1}  \bigg[ \frac{ \mathrm{PAV} - \mathrm{PAV_{low}}}{\mathrm{PAV}} \bigg]_i ,
    \label{eqn:dpcpeqn}
\end{align}
where $\mathrm{PAV}_i$ is the ring-averaged total pressure and $\mathrm{PAV_{low, \, i}}$ is the ring-averaged pressure in the low pressure region (the azimuthal extent of the ring where the pressure falls below $\mathrm{PAV}_i$).\\

Fig.~\ref{Fig11}(a) presents the impact of the suction on these quantities. The suction amplitude represents the vertical velocity normalized by the incoming speed of sound. For low amplitudes $u_z/c_{\infty} < 0.1$, the pressure recovery (PR) is lower than that of the unforced (natural) case. At low amplitudes, the forcing is insufficient at the second bend. Additionally, the forcing at the first bend pulls fluid downward, counteracting the suction at the second bend and thereby increasing the intensity of vortices at the second bend, which results in increased distortion. As the suction amplitude increases, PR initially improves, reaching 0.992 (a 0.65\% increase) at $u_z/c_{\infty}=0.15$, and decreases thereafter. Physically, at those high amplitudes, although the streamwise vortices near the upper wall are weakened, those near the lower wall are strengthened, reducing the total pressure near the lower wall and thereby lowering the overall pressure recovery. The temporally averaged flow distortion, $\langle \rm{dPcP} \rangle$, reduces with increasing suction.  The case corresponding to the highest $\rm{PR}$ also has a reduced mean distortion, $\langle \rm{dPcP} \rangle$, compared to the unforced flow. Thus, among the cases considered here, $u_z/c_{\infty}=0.15$ is considered an optimal suction condition. \\

\setcounter{table}{0}
\begin{table}
\centering
\begin{tabular}{ p{1.5cm}p{4cm}p{4.6cm}}
\hline\hline
Case & First bend amplitude & Second bend amplitude  \\
\hline
OC-1    & $u_z(\Delta z/D_{AIP} \approx 0.1)/c_{\infty}$ & $u_z(\Delta z/D_{AIP} \approx 0.1)/c_{\infty}$   \\
OC-2    & $u_z(\Delta z/D_{AIP} \approx 0.1)/c_{\infty}$ & $u_z(\Delta z/D_{AIP} \approx 0.1)/c_{\infty} +0.025$   \\
OC-3    & $u_z(\Delta z/D_{AIP} \approx 0.1)/c_{\infty}$ & $u_z(\Delta z/D_{AIP} \approx 0.1)/c_{\infty} +0.05$   \\
OC-4    & $u_z(\Delta z/D_{AIP} \approx 0.1)/c_{\infty}$ & $u_z(\Delta z/D_{AIP} \approx 0.1)/c_{\infty} +0.1$   \\
\hline\hline
\end{tabular}
\caption{Control amplitude for different opposition control cases, normalized by the speed of sound $c_{\infty}$. Here, $\Delta z$ represents the wall-normal distance from the wall surface, with $\Delta z = z_{\rm{wall}} - z$ at the first bend (bottom wall) and $\Delta z = z - z_{\rm{wall}}$ at the second bend (top wall). }
\label{table:OC-cases}
\end{table} 

Next, the effect of different suction amplitudes at different hole locations is assessed. To this end, the suction amplitude at each row of holes is chosen to match the mean vertical velocity measured at a wall-normal distance of $ z/D_{AIP} \approx 0.1$ in the unperturbed flow. We note that this choice of location is ad hoc, but distances closer to the wall than this did not yield attenuation of the streamwise vortices. This approach represents a transpiration-based opposition control, referred to as case OC-1 in Fig.~\ref{Fig11}(b). However, this case yields a worse pressure recovery and time-averaged flow distortion at the AIP in comparison to the unforced flow. Subsequently, a parametric study exploring suction amplitudes within a narrow band around the OC-1 values, ranging from 0.9 to $1.1 \times u_z(\Delta z/D_{AIP} \approx 0.1)/c_{\infty}$ was conducted. Here, $\Delta z$ represents the wall-normal distance from the wall surface, with $\Delta z = z_{\rm{wall}} - z$ at the first bend (bottom wall)  and $\Delta z = z - z_{\rm{wall}}$ at the second bend (top wall). Due to the large parameter space, only configurations demonstrating performance improvements are reported, with key trends summarized as follows. For the first bend (bottom wall), reducing the suction amplitude below the OC-1 value degraded performance whereas only a slight increase proved beneficial. For the second bend (top wall), reducing the amplitude below $u_z(\Delta z/D_{AIP} \approx 0.1)/c_{\infty}$ negatively affected the flow characteristics, whereas an increase in amplitude led to improved performance. To that end, Fig.~\ref{Fig11}(b) presents three optimized cases in addition to the transpiration-based opposition control (OC-1). Cases (OC-2) -- (OC-4) employ the same amplitude of $u_z(\Delta z/D_{AIP} \approx 0.1)/c_{\infty}$ at the bottom wall; but different amplitudes on the top wall. Case OC-3 provides the most favorable performance, characterized by a 0.85\% increase in pressure recovery and a substantial (60\%) reduction in time-averaged flow distortion. \\

\begin{figure}[!ht]
\centering
{\includegraphics[trim={0.0cm 3cm 0cm 1.8cm},clip,width=1.0\textwidth]{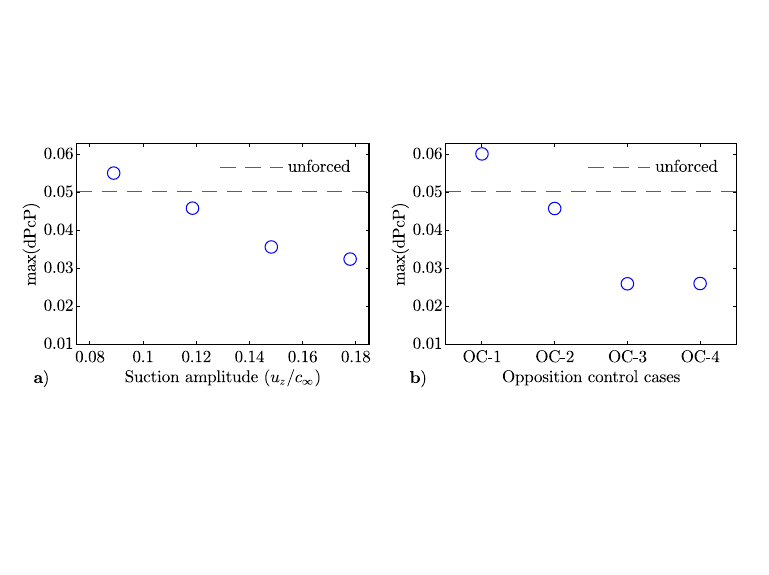}}
      	\caption{ Effect of ($\textbf{a}$) suction amplitude and ($\textbf{b}$) different opposition control cases, reported in table \ref{table:OC-cases}, on maximum flow distortion, $ \rm{max(dPcP)}$. The unperturbed case is denoted by dashed lines.} \label{Fig11b}
\end{figure}

Fig.~\ref{Fig11b} presents the predicted maximum flow distortion, $\mathrm{max(dPcP)}$, for the different controlled cases. Similar trends in the maximum and time-averaged flow distortion are evident. For uniform suction amplitude across all holes, increasing the amplitude decreases the maximum distortion. The lowest maximum distortion with uniform suction is 0.0325 ($\approx 36\%$ reduction from the baseline), achieved at $u_z/c_{\infty} = 0.18$. The optimized opposition control cases achieve even greater reductions, with OC-3 attaining the lowest value of 0.026, approximately 49\% reduction from the unforced case. These results demonstrate that the opposition control strategies employed in this study effectively enhance pressure recovery while simultaneously reducing both the time-averaged and maximum flow distortion.

\begin{figure}[!ht]
            {\includegraphics[trim={0.0cm 3.1cm 0cm 3.2cm},clip,width=1.0\textwidth]{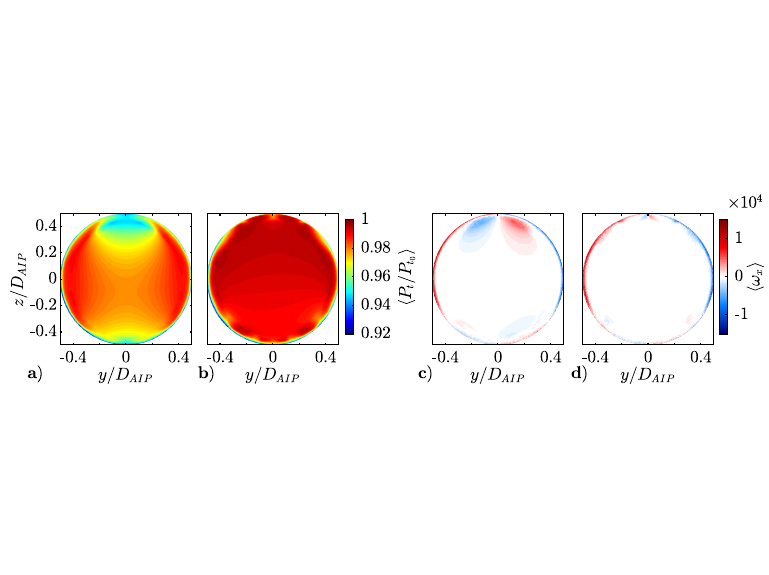}}
      	\caption{Temporally averaged total pressure fields (a,b) and mean streamwise vorticity (c,d) at the AIP for the natural (a,c) and controlled (b,d) cases. The controlled case corresponds to OC-3.}\label{Fig12} 
\end{figure}

To further elucidate the effect of flow control, Fig.~\ref{Fig12} compares the mean total pressure and mean streamwise vorticity between the unforced case and the optimally controlled case, OC-3, in the AIP for $Ma_{AIP}=0.36$ flow. For the controlled case, the pressure recovery improves near the top and bottom walls, likely as the transpiration-based opposition control removes low-momentum fluid from these wall regions, reducing the low-pressure zones present in the unforced flow. The mean streamwise vorticity, shown in Fig.~\ref{Fig12}(c,d), reveals that the opposition control suppresses the quasi-steady streamwise vortices, particularly at the top wall. This suppression reduces the advection of low-speed fluid from the near-wall region into the core. \\

\subsection{Power-Efficiency analysis}
Next, as a practical measure, an input-output power analysis is performed to characterize the efficiency of the proposed control methodology. The power input is estimated as:
\begin{equation}
\mathcal{P}_{\text{input}} = \frac{1}{2}\rho v^3 A_{\text{holes}}n_{\text{holes}}
\end{equation}
where $n_{\text{holes}}$, $A_{\text{holes}}$, and $v$ are the number of holes, area of each hole, and amplitude of suction velocity in each hole, respectively. The net power gain due to control is defined as:
\begin{equation}
\mathcal{P}_{\text{gain}} = \bqty{\langle P_{t_i} u_i\rangle_{i}  {A_{\text{AIP}}}}_{\text{controlled}} - \bqty{\langle P_{t_i} u_i\rangle_{i}  {A_{\text{AIP}}}}_{\text{uncontrolled}}.
\end{equation}
where $A_{\text{AIP}}$ is the area of the AIP plane, the subscript $i$ refers to the $i$-th probe location, $u_i$ is the streamwise velocity at the $i$-th probe, $P_{t_i}$ is the total pressure at the $i$-th probe, and $\langle\cdot\rangle_{i}$ denotes temporal averaging followed by spatial averaging over all 40 probes at the AIP. The control efficiency is defined as
\begin{equation}
\eta (\%) = \frac{\mathcal{P}_{\text{gain}} - \mathcal{P}_{\text{input}}}{\mathcal{P}_{\text{input}}} \times 100.
\end{equation}
Positive values of $\eta$ indicate that the control strategy is favorable from a power input perspective. Table \ref{table:power_analysis} reports the power input, power gain, and control efficiency for different forcing configurations. Only cases that improve pressure recovery are reported. It is observed that OC-3, OC-4, and cases with suction amplitude ($A  \geq 0.15$) across all holes are power efficient. Two cases are noteworthy: the most power efficient case corresponds to the suction amplitude  case, with $A = 0.18$; however, it results in a lower pressure recovery than other controlled cases. On the other hand, Case OC-3 achieves the highest pressure recovery and lowest flow distortion with an efficiency of approximately $65\%$.

\begin{table}[!ht]
\centering
\begin{tabular}{ p{2.8cm}p{2.5cm}p{2.4cm}p{2cm}p{2cm}p{2.7cm}}
\hline\hline
Case & Power input (KW) & Power gain (KW) & Efficiency ($\eta$) & Gain (G) &   $\Delta\text{PR}$  \\
\hline
Suction  $A=0.12$  & 0.67 & 0.38 & -42.5\% & 0.29\% & 0.0033    \\
Suction  $A=0.15$  & 1.31 & 2.05 & 57.4\% & 1.55\%& 0.0065    \\
Suction  $A=0.18$ & 2.25 & 4.76  & 111.6 \% & 3.58\% & 0.0052   \\
OC-2  & 1.32 & 0.90 & -32.3\% & 0.68\% & 0.0058    \\
OC-3 & 1.63 & 2.69 & 64.7\% & 2.03\% & 0.0087    \\
OC-4  & 1.86 & 2.48 & 33.5\% & 1.87\% & 0.0075  \\
\hline\hline
\end{tabular}
\caption{Power input, power gain, control efficiency, and gain for different forcing configurations. The increase in pressure recovery relative to the uncontrolled baseline (with pressure recovery of 0.986) is shown for comparison. Note that the suction amplitude, $A = u_z (\Delta z/D_{AIP} \approx 0.1)$.  }
\label{table:power_analysis}
\end{table} 

Another measure of the gain relative to the total input (input of control and the power at the inlet) is defined as,  
\begin{equation}
G = \frac{\mathcal{P}_{\text{gain}}}{\mathcal{P}_{\text{input}}+\mathcal{P}_{\text{inlet}}} \times 100.
\end{equation}
This metric quantifies the net increase in power at the AIP relative to the total power input to the system. The gain for the different control cases are reported in Table \ref{table:power_analysis}; the uniform suction case with $A=0.18$ offers the highest gain of 3.58\%, while the OC-3 case offers a gain of 2.03\%.  The optimized steady opposition control thus provides a net positive power gain alongside enhanced pressure recovery and a reduced flow distortion.

\section{Effect of Mach Number on opposition control viability} \label{sec:M_effects}

In this section, the efficacy of the proposed opposition control strategy with increasing Mach number is examined. Although not shown, applying constant amplitude suction at higher Mach numbers ($Ma_{AIP} = 0.46$ and 0.54) produced adverse outcomes compared to their respective unforced baseline flows. This is because an increase in Mach number intensifies the streamwise vortices and subsequently broadens the range of wall-normal velocities across the bends, thus making a constant (and equal) suction across all holes less effective. For example, at $Ma_{AIP} = 0.36$,  the mean vertical velocity componentat the first and second bends exhibits similar magnitudes, ranging between $0.1 \leq |u_z/c_{\infty}| \leq 0.175$. In contrast, at $Ma_{AIP} = 0.54$, the mean wall-normal velocity at the first bend reaches approximately  $|u_z/c_{\infty}| \approx 0.45$, whereas at the second bend it is about $|u_z/c_{\infty}| \approx 0.15$. Consequently, only the results based on opposition control strategy, which yields favorable results, are discussed hereafter.

\subsection{Pressure recovery and distortion metrics}

\begin{figure}[!ht]
\centering
{\includegraphics[trim={0.0cm 3cm 0cm 2.7cm},clip,width=1.0\textwidth]{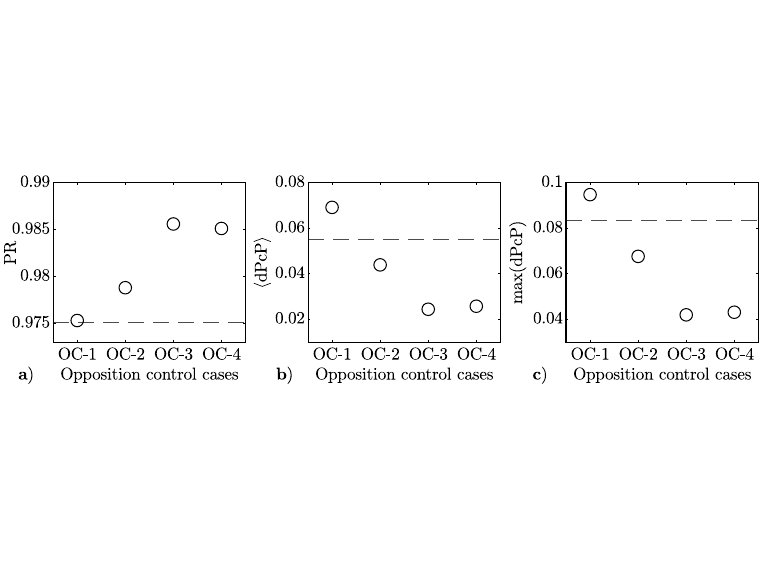}}
      	\caption{ Effect of different opposition control cases for $Ma_{AIP}=0.46$, on (a) pressure recovery, (b) time averaged flow distortion, and (c) maximum flow distortion. The unperturbed case is denoted by dashed lines.} \label{Fig14}
\end{figure}

Opposition control is applied at $Ma_{AIP}=0.46$ using amplitude estimates similar to those for $Ma_{AIP}=0.36$. Fig.~\ref{Fig14} shows the pressure recovery, time-averaged flow distortion, and maximum flow distortion for four opposition control cases. The trends are similar to those observed in Fig.~\ref{Fig11}. OC-1 provides no improvement in pressure recovery, while OC-2, OC-3 and OC-4 increase pressure recovery and reduce flow distortions. In particular, case OC-3 exhibits the highest pressure recovery and lowest flow distortion at this intermediate Mach number flow. Specifically, OC-3 reduces the time-averaged flow distortion by 55\% and maximum flow distortion by 50\% relative to the unforced case. \\ 

\begin{table}
\centering
\begin{tabular}{ p{1.8cm}p{0.8cm}p{1.45cm}p{1.32cm}p{1.25cm}p{1.25cm}p{1.35cm}p{1.47cm}p{1.45cm}}
\hline\hline
Case name & OC-1 & OC-2 & OC-3 & OC-4 & OC-5 & OC-6 & OC-7 & OC-8  \\
\hline
First bend amplitude    & $A_0$ & $A_0$  & $A_0$ & $A_0$ & $A_0-0.1$ & $A_0-0.05$  & $A_0-0.025$ & $A_0-0.025$   \\
Second bend amplitude    & $A_0$ & $A_0+0.025$ & $A_0+0.05$ & $A_0+0.1$ & $A_0$ & $A_0$ & $A_0$ & $A_0+0.075$   \\
$\Delta \mathrm{PR}$ & $-0.011$ & $-0.011$ & $-0.0045$ & $-0.0084$ & $-0.010$ & $-0.0056$ & $-0.012$ & $-0.0037$  \\
\hline\hline
\end{tabular}
\\
\vspace{0.1cm}
\textbf{$\textbf{a})$ Cases 1 -- 8}

\vspace{0.5cm} 
\begin{tabular}{ p{1.8cm}p{1.35cm}p{1.42cm}p{1.42cm}p{1.42cm}p{1.42cm}p{1.42cm}p{1.42cm}}
\hline\hline
Case name & OC-9 & OC-10 & OC-11 & OC-12 & OC-13 & OC-14 & OC-15 \\
\hline
First bend amplitude    & $A_0-0.05$  & $A_0-0.05$ & $A_0-0.055$ & $A_0-0.055$& $A_0-0.065$&  $A_0-0.065$  & $A_0-0.065$   \\
Second bend amplitude    & $A_0+0.05$  & $A_0+0.055$ & $A_0+0.06$ &$A_0+0.065$ & $A_0+0.065$ & $A_0+0.075$ & $A_0+0.08$   \\
$\Delta \mathrm{PR}$ & $-0.0003$ & $0.0007$ & $0.0019$ & $0.0017$ & $0.0026$ & $-0.0008$  & $-0.0003$  \\
\hline\hline
\end{tabular}
\\ 
\vspace{0.1cm}
\textbf{$\textbf{b})$ Cases 9 -- 15}
\caption{Control amplitude and change in pressure recovery for opposition control cases at $Ma_{AIP}=0.54$. Amplitudes are normalized by the freestream speed of sound $c_{\infty}$, where $A_0=u_z(z_s)$ and $z_s=\Delta z/D_{\text{AIP}}$ represents the wall-normal sensing location. At the first bend (bottom wall), $z_s = z_{\rm{wall}} - z$; at the second bend (top wall), $z_s = z - z_{\rm{wall}}$.}
\label{table:OC-cases_M0p54}
\end{table}

\begin{figure}[!ht]
            {\includegraphics[trim={0.0cm 3.1cm 0cm 1.4cm},clip,width=1.0\textwidth]{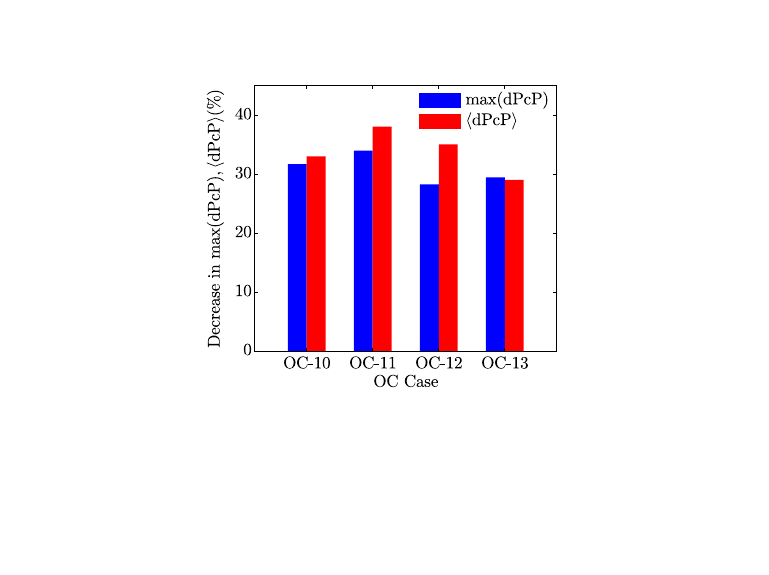}}
            \caption{Effect of opposition control on flow distortion for $Ma_{AIP}=0.54$. The percentage decrease in time-averaged and maximum flow distortion for the different control cases is shown. }
\label{Fig16}
\end{figure}

\noindent
Next, we follow a similar procedure to tune the control amplitude for $Ma_{AIP}=0.54$. At this Mach number, a weak upstream shock exists, and the proposed flow strategy required tuning in a narrower range of amplitudes to increase total-pressure recovery. The opposition-control cases with amplitudes similar to those used for $Ma_{AIP}=0.36$ and $0.46$ resulted in a decrease in total-pressure recovery. Table \ref{table:OC-cases_M0p54} summarizes 15 cases, listing the control amplitude applied at each bend and the corresponding net change in pressure recovery. Among these cases, only OC-10, OC-11, OC-12, and OC-13 yield an increase in total-pressure recovery, although the improvement is small. Nevertheless, as demonstrated in Fig.~\ref{Fig16}, both the time-averaged and maximum distortion are reduced. The reductions range from approximately $28\%$ to $38\%$ relative to the uncontrolled case, indicating that opposition control remains effective in mitigating distortion despite the small gains in total-pressure recovery.  Although no systematic trend is observed, the results suggest that a slight reduction in control amplitude relative to the sensing amplitude at the first bend, combined with a slight increase in amplitude at the second bend, improves pressure recovery. In particular, case OC-11 provides the largest overall reduction in time-averaged and maximum flow distortion.

\subsection{Power efficiency analysis at higher Mach numbers}

Table \ref{table:power_analysis_Ma} presents the efficiency of the optimized control cases, i.e., the cases that produce the highest pressure recovery and the lowest flow distortion. For both these Mach numbers, $Ma_{AIP}=0.46$ and 0.54, the required control power exceeds the power recovered, resulting in an efficiency of $\eta \leq 0$. In contrast to the results at $Ma_{AIP}=0.36$ reported in Table \ref{table:power_analysis}, these control cases are therefore not power-efficient. This reduced efficiency is primarily attributed to the larger wall-normal velocity fluctuations at higher Mach numbers, particularly near the flow-separation region. Consequently, opposition control requires larger actuation amplitudes and, thus, greater power input. Nevertheless, both cases yield a small net increase in the overall power balance $\leq 1\%$ relative to the combined inlet and control-input power. Thus, although the energetic benefit is modest at the higher Mach numbers, the opposition-control strategy still provides a net positive power gain while simultaneously improving pressure recovery and reducing flow distortion.

\begin{table}
\centering
\begin{tabular}{ p{1.1cm}p{1.1cm}p{2.6cm}p{2.4cm}p{2cm}p{1.3cm}p{1.2cm}}
\hline\hline
\centering Case & \centering $Ma_{AIP}$ & \centering Power input (KW) & \centering Power gain (KW) & \centering Efficiency ($\eta$) & \centering Gain (G) &   $\Delta\text{PR}$  \\
\hline
\centering OC-3   &\centering 0.46 & \centering 2.95 & \centering 0.88 & \centering-70.1\% & \centering 0.55\% & 0.0111  \\
\centering OC-11  & \centering 0.54  &  \centering 3.74 & \centering 1.36 & \centering -63.6\% & \centering 0.76\%& 0.0016    \\
\hline\hline
\end{tabular}
\caption{Power input, power gain, control efficiency, and gain for different forcing configurations. The increase in pressure recovery relative to the uncontrolled baseline (with pressure recovery of 0.986) is shown for comparison.}
\label{table:power_analysis_Ma}
\end{table} 


\subsection{Controlled flow patterns }
\begin{figure}[!ht]
            {\includegraphics[trim={0.0cm 2.7cm 0cm 0.8cm},clip,width=1.0\textwidth]{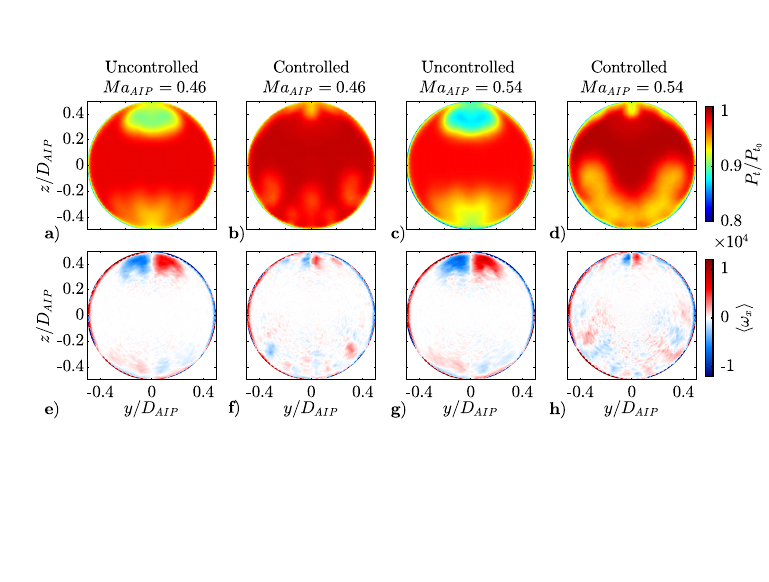}}
        \caption{Temporally averaged total pressure fields (a$-$d) and mean streamwise vorticity (e$-$h) at the AIP for the uncontrolled and controlled cases: (a,e) uncontrolled and (b,f) controlled at $Ma_{AIP}=0.46$; (c,g) uncontrolled and (d,h) controlled at $Ma_{AIP}=0.54$. The controlled cases correspond to OC-3 and OC-11 for $Ma_{AIP}=0.46$ and $0.54$, respectively.}\label{Fig17}
\end{figure}

Beyond the quantitative distortion metrics, the flow fields of the controlled cases can be compared with baseline cases. To that end, Fig.~\ref{Fig17} shows total-pressure and streamwise-vorticity fields are shown in the top and bottom rows, respectively. Figs.~\ref{Fig17}(a,e) and (c,g) correspond to the unforced cases at $Ma_{AIP}=0.46$ and $0.54$, respectively, while figures \ref{Fig17}(b,f) and (d,h) show their controlled counterparts. For $Ma_{AIP}=0.46$, opposition control diminishes the regions of low total pressure near both the top and bottom walls. This improvement is associated with the suppression of the streamwise vortices, particularly near the top wall, as evident from the comparison of panels (e) and (f). The weakened vortices reduce the advection of low-momentum fluid away from the wall and consequently improve the total pressure. Near the bottom wall, the large streamwise vortices observed in the unforced case break down into smaller structures and are redistributed away from the centerline in the azimuthal direction. As a result, the concentrated region of low total pressure near the centerline is weakened and distributed over a broader azimuthal extent in the controlled case. Similar but more pronounced, changes are observed at the higher Mach number, $Ma_{AIP}=0.54$. The low total-pressure region near the top wall is substantially reduced, while that near the bottom wall becomes less intense and is distributed over a broader region. Consistent with these changes, the streamwise vortices near the top wall are considerably weakened, whereas the large vortical structures near the bottom wall break down into smaller structures and spread over a broader azimuthal extent.

\begin{figure}[!ht]
            {\includegraphics[trim={0.0cm 3.5cm 0cm 2cm},clip,width=1.0\textwidth]{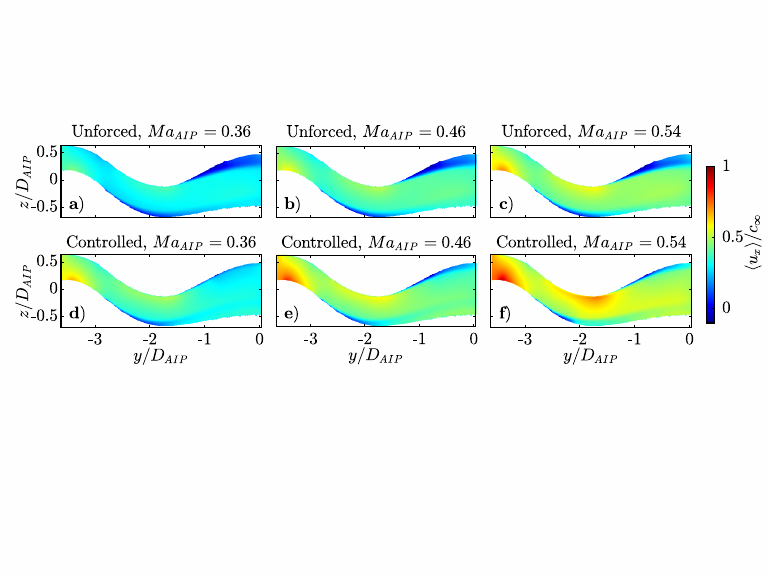}}
        \caption{Mean streamwise velocity on the $x$-$z$ plane at $y=0$, for the uncontrolled cases at (a) $Ma_{AIP}=0.36$, (b) $Ma_{AIP}=0.46$ and (c) $Ma_{AIP}=0.54$, and the corresponding controlled cases at (d) $Ma_{AIP}=0.36$, (e) $Ma_{AIP}=0.46$ and (f) $Ma_{AIP}=0.54$. }\label{Fig18}
\end{figure}

\noindent
We now investigate the effect of control on flow separation. Fig.~\ref{Fig18} presents a side-view of the mean streamwise velocity for the unforced (a-c) and controlled (d-f) cases. The uncontrolled cases exhibit low-velocity regions along the lower and upper walls downstream of the first and second bends, respectively, indicating flow reversal and separation. Opposition control weakens these velocity deficits and reduces the extent of the separated regions at both bends for all three Mach numbers considered. The imposed forcing acts as a suction-like control, removing low-momentum fluid from the near-wall region and thereby reducing flow separation. Consequently, the separated shear layer lies closer to the wall. This behavior is consistent with previous studies demonstrating the effectiveness of suction in suppressing flow separation and promoting reattachment \cite{greenblatt2006experimental,Amitay2002,chen2013flow}. For example, Greenblatt \emph{et al.} \cite{greenblatt2006experimental} showed that suction accelerates the flow upstream of the actuation slot and increases the pressure downstream of the slot, thereby shortening the separation bubble. The reduction in separation observed here is therefore consistent with the established mechanism of suction-based flow control. The upstream acceleration induced by the suction also explains the increase in mean streamwise velocity observed from the inlet to the first bend in the controlled cases across all three Mach numbers.

\subsection{Control effects on pressure fluctuations}
\begin{figure}[!ht]
 \centering
            {\includegraphics[trim={0.0cm 0cm 0cm 0cm},clip,width=1.0\textwidth]{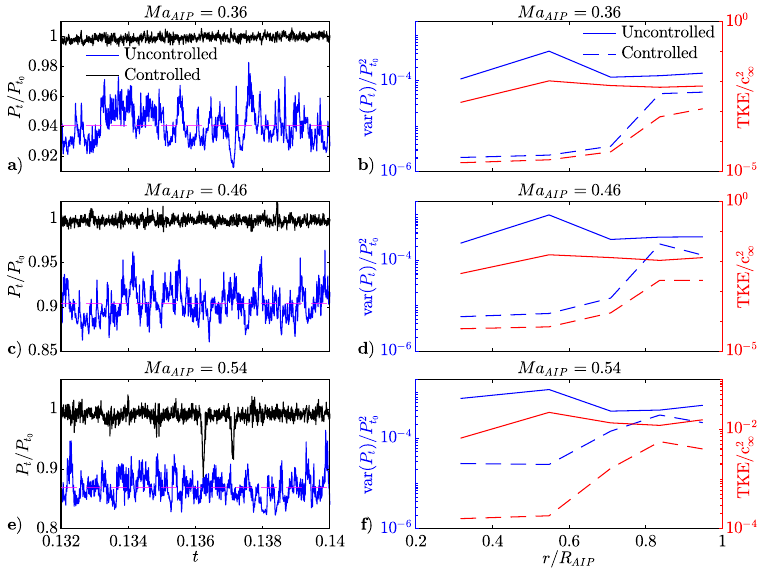}}
        \caption{Effect of opposition control on pressure fluctuations and TKE: (a,c,e) total-pressure histories at $(y/R_{AIP},z/R_{AIP})=(0,0.71)$; (b,d,f) total-pressure variance (blue, left ordinate) and TKE (red, right ordinate) along Rake 1 ($y=0$, $0.31 \leq z/R_{AIP} \leq 0.95$). Controlled cases correspond to OC-3 for $Ma_{AIP}=0.36,0.46$ and OC-11 for $Ma_{AIP}=0.54$. In (b), solid and dashed lines denote uncontrolled and controlled cases, respectively; only blue curves are labeled.}\label{Fig19}
\end{figure}

Finally, the effect of steady opposition control on fluctuating quantities is examined in Fig.~\ref{Fig19}. The total-pressure time histories on one of the pressure probes on the AIP at $(y/R_{AIP},z/R_{AIP})=(0,0.71)$ are shown in Fig.~\ref{Fig19}(a,c,e) for the unforced and controlled cases at the three Mach numbers considered. A short but representative time window is shown. The unforced case exhibits large fluctuations about the mean (dashed magenta line). In contrast, controlled cases enhance the total pressure around $P_t/P_{t_0}\approx 1$ across all mach numbers, and the time history maintains a more uniform total pressure with substantially suppressed fluctuations, especially for $Ma_{AIP}=0.36,0.46$. For the highest Mach number $Ma_{AIP}=0.54$, the opposition control also attenuates the pressure fluctuations, but does not completely suppress the intermittent events associated with the largest instantaneous pressure losses. 

For a quantitative comparison, Fig.~\ref{Fig19}(b,d,f) shows the variance of total pressure and turbulent kinetic energy along Rake 1 at the AIP, which extends vertically from the center to the top wall, i.e. $y/R_{AIP}=0$ $0.3 \leq z/R_{AIP} \leq 0.95$. The probes along the Rake 1 are chosen because they lie in the region of lowest total pressure in the uncontrolled flow, which contributes most strongly to the distortion. The total pressure variance is shown on the left ordinate, while TKE is shown on the right ordinate, with the uncontrolled and controlled cases denoted by solid and dashed lines, respectively. The controlled cases exhibit significant reductions in both quantities, with similar trends across all three Mach numbers, although the reductions become slightly smaller as the Mach number increases.  Closer to the center of the plane ($z/R_{AIP} \leq 0.5$), control reduces both quantities by approximately two orders of magnitude, with the reduction gradually decreasing toward the periphery to a factor of three at $z/R_{AIP} = 0.95$. Overall, these results show that, in addition to the reduction in mean and maximum flow distortion, opposition control is able to reduce pressure fluctuations and turbulent kinetic energy.


\section{Concluding Remarks} \label{conclusions}

This study investigates the flow physics in a serpentine duct (SD-2 of \citet{burrows2021experimental}), characterizing the Gortler vortices that lead to flow distortion and loss of total pressure, and performs steady opposition control to attenuate these vortices and improve pressure recovery. Following Agrawal et al. \cite{agrawal2025predictions},  wall-modeled large eddy simulations are conducted at three Mach numbers, $Ma_{AIP} = 0.36, \, 0.46$, and $0.54$. The duct contains two longitudinal bends and its cross-section transitions from a D-shape at the inlet to a circular shape at the AIP. These geometric features introduce adverse pressure gradients and centrifugal forces that induce flow separation, producing a pair of counter-rotating Görtler vortices. As the flow evolves downstream, these vortices are advected from the periphery toward the duct centerline along the spanwise direction. The Görtler vortices eject low-momentum fluid from near the wall into the core, leading to azimuthal distortion and a loss of pressure recovery. This phenomenon is observed consistently across all  Mach numbers considered herein. The vortices formed near the top wall are stronger than those near the bottom wall, causing greater total pressure loss in the upper region of the duct. 

Given these flow recovery losses, steady opposition control was pursued to suppress of the streamwise vortices. Twelve holes were introduced at each bend, arranged in three rows of four holes each, in the region of flow separation, with the first row located at the onset of flow separation. Opposition control based on vertical velocity was implemented as a boundary condition at these holes. Two control strategies were implemented: constant amplitude suction across all holes, and variable amplitude suction based on the mean vertical velocity measured at $ \Delta z/D_{\rm{AIP}} \approx 0.1$ from the wall. For both strategies, a parametric study was conducted to identify the optimal amplitude that maximizes pressure recovery. At the second bend, an amplitude slightly higher than the velocity at the sensing location is needed to increase pressure recovery and reduce distortions.  This control strategy is applied, and shown to reduce flow distortions across all three Mach numbers considered. 
The increases in pressure recovery and reductions in time-averaged and maximum flow distortion are $0.0085$, $60\%$, and $49\%$, respectively, for $Ma_{AIP}=0.36$; $0.011$, $55\%$, and $50\%$ for $Ma_{AIP}=0.46$; and $0.0019$, $38\%$, and $34\%$ for $Ma_{AIP}=0.54$.  Although the higher Mach number case is more challenging and exhibits a narrower range of effective control amplitudes, opposition control still yields improvements, particularly in attenuating the mean as well as maximum flow distortion. The physical mechanism behind this improvement is that opposition control attenuates the Görtler vortices, thereby limiting the advection of low-speed fluid from the wall to the core and enhancing pressure recovery.

An input-output power analysis was performed to assess the energetic efficiency of the steady opposition-control strategy. At $Ma_{AIP}=0.36$, the recovered power exceeded the control input, yielding a maximum efficiency of approximately $\eta=65\%$. In contrast, for $Ma_{AIP}=0.46$ and $0.54$, the net power gain remained positive but was smaller than the required control input, resulting in negative values of the defined efficiency. The reduced efficiency at the higher Mach numbers is attributed to the larger actuation amplitudes required to achieve effective control. These results indicate that, while opposition control remains aerodynamically beneficial across all Mach numbers, its energetic efficiency decreases as the Mach number increases. Nevertheless, a major benefit of control is that it substantially suppresses the total-pressure fluctuations across all Mach numbers. In particular, the variance of the total pressure is reduced by approximately two orders of magnitude near the center of the AIP and by about a factor of two closer to the upper edge. The magnitudes of the Reynolds-stress events are also reduced, especially the ejection events, which are associated with the time instances of the largest total pressure losses. Such extreme events can adversely affect compressor operability and increase the likelihood of stall; therefore, their suppression is particularly important and further demonstrates the effectiveness of the opposition-control strategy.


Overall, it is noted that the decreasing control efficiency at higher Mach numbers may highlight potential limitations of active opposition control at flight-relevant Mach numbers and motivate the development and application of other approaches, such as geometric shape optimization to attempt reducing flow distortions in serpentine diffusers.

\section*{Acknowledgments}
The support of ONR to CTR under grant N000142312833 is gratefully acknowledged.  This research used resources of the Oak Ridge Leadership Computing Facility, which is a DOE Office of Science User Facility supported under Contract DE-AC05-00OR22725. The authors acknowledge fruitful discussions with Dr. Chad Winkler from The Boeing Company. 

\section*{Appendix I: Quadrant analysis} \label{appendix:A}

In this section, we examine which Reynolds-stress motions, particularly ejection and sweep events, are associated with time instances corresponding to extrema in the total-pressure fluctuations. Recall that the streamwise velocity and vorticity contours in Figs~\ref{Fig2} and \ref{Fig3} suggested that along the spatial extent of the AIP, the low-momentum fluids gets advected from the near-wall region toward the core. To gain further insight into this process, a quadrant analysis is performed near the top wall on the AIP at $(x/D_{AIP},y/D_{AIP},z/D_{AIP})=(0,0,0.355)$. Herein, the Reynolds shear stress ($-u'w'$ where $u'$ and $w'$ are the streamwise and wall-normal velocity fluctuations) is decomposed into four quadrants: $\mathrm{Q1}$, $\mathrm{Q2}$, $\mathrm{Q3}$, and $\mathrm{Q4}$. In quadrant $\mathrm{Q1}$, $u' > 0$ and $w' > 0$; in $\mathrm{Q2}$, $u' < 0$ and $w' > 0$; in $\mathrm{Q3}$, $u' < 0$ and $w' < 0$; and in $\mathrm{Q4}$, $u' > 0$ and $w' < 0$. The $\mathrm{Q2}$ and $\mathrm{Q4}$ motions correspond to ejection and sweep events, respectively, whereas $\mathrm{Q1}$ and $\mathrm{Q3}$ represent outward and inward interactions \citep{corino1969visual, wallace1972wall}. \\

\begin{figure}[!t]
\centering
            {\includegraphics[trim={0.0cm 2.2cm 0cm 0cm},clip,width=1\textwidth]{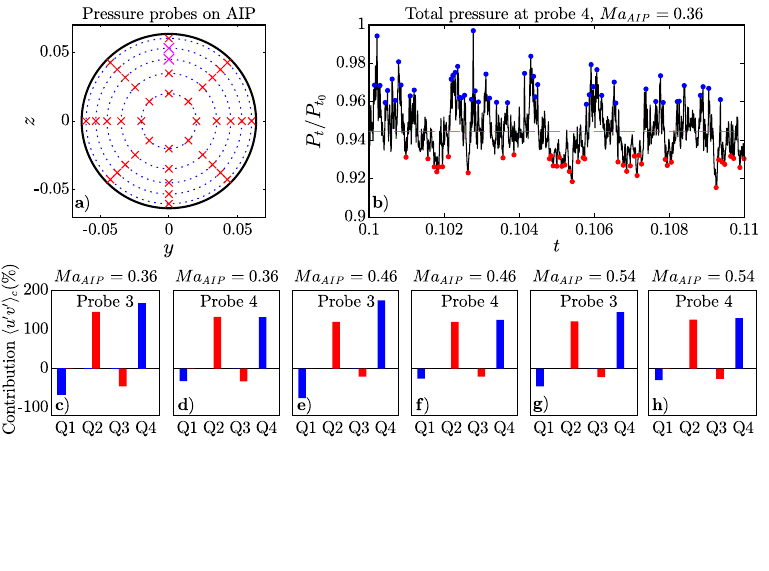}}
      	\caption{ (a) Pressure probes distribution on the AIP, where $y$ and $z$ correspond to the spanwise and the vertical directions, respectively. (b) Time history of total pressure at  Probe 4 ($y/D_{AIP}=0$, $z/D_{AIP}=0.4$). A representative time interval is shown with local maxima and minima denoted by blue and red circles, respectively. Panels (c)-($h$) show the percentage contribution of quadrant events $\mathrm{Q1}$, $\mathrm{Q2}$, $\mathrm{Q3}$, and $\mathrm{Q4}$ to Reynolds shear stress conditioned on the local maxima and minima of total pressure for $Ma_{\rm AIP} = 0.36, 0.46,$ and $0.54$. \label{Fig6} }
\end{figure}

Further, to correlate the contribution of the quadrant events to the local minima and maxima of total pressure at the AIP, particularly along the centerspan region on the top wall. For reference, Fig.~\ref{Fig6}(a) presents the distribution of 40 dynamic pressure probes at the AIP.  The probes positioned along $y/D_{AIP} = 0$ and $z/D_{AIP} > 0$ are located within the region of intense pressure losses, with the third and fourth probes (marked by $\times$ in magenta) placed closest to the point of maximum total pressure loss. The temporal history of the total pressure at the fourth probe for $Ma_{AIP}=0.36$ flow is shown in Fig.~\ref{Fig6}(b). For visual clarity, only a small representative interval is presented. The mean total pressure at this probe is denoted by the dashed magenta line, while the local maxima and minima of the time series are marked by blue and red circles, respectively. The 500 most prominent maxima and minima are identified over the time history, where the minima represent instances of high pressure losses and vice-versa.  
\\


The contributions of the four quadrant events at the time instances corresponding to the local maxima and minima of total pressure are presented in Fig.~\ref{Fig6}(c)-(h) on probes 3 and 4. Events conditioned on local maxima and minima are indicated in blue and red respectively. The contributions of all four events sum to 100\%. If a particular event exceeds 100\%, it indicates that the other events contribute negatively, opposing its effect. A common trend is observed across both probes and all Mach numbers: only $\mathrm{Q2}$ and $\mathrm{Q3}$ events contribute during intense pressure losses, whereas $\mathrm{Q1}$ and $\mathrm{Q4}$ events contribute to instances of higher pressure recovery. This further supports the conjecture that the ejection events drive the loss of total pressure (which also enhances azimuthal distortion), whereas the sweep events are associated with pressure gains (and reduce azimuthal distortions). The inward and outward interactions contribute negatively to the Reynolds shear stress, indicating that they counteract the effects of the ejection and sweep events during maximum and minimum azimuthal distortions. Similar analyses are extended to flow resulting from controlled actuation and presented below.

\begin{figure}[!ht]
 \centering
            {\includegraphics[trim={0.0cm 0cm 0cm 0cm},clip,width=1.0\textwidth]{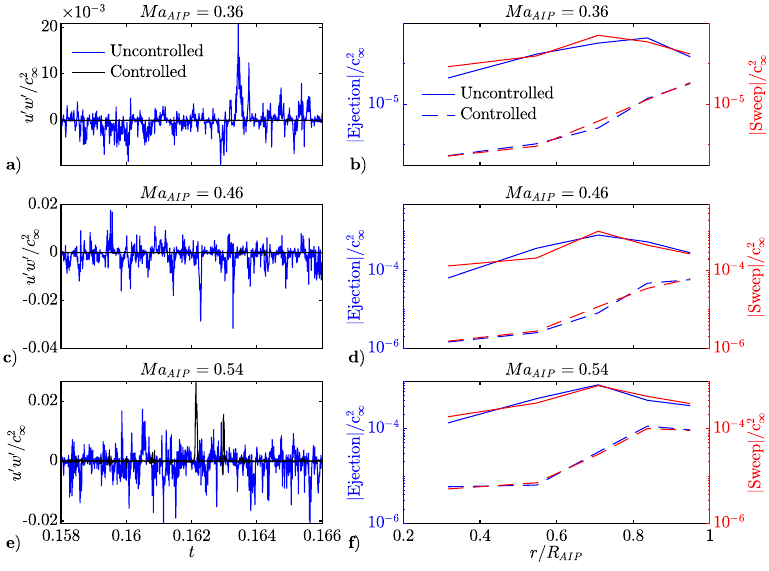}}
            \caption{Effect of opposition control on Reynolds stress events: (a,c,e) time histories of $u'w'$ at $(y/R_{AIP},z/R_{AIP})=(0,0.71)$ on the AIP; (b,d,f) magnitudes of ejection (blue, left ordinate) and sweep events (red, right ordinate) along Rake 1 ($y=0$, $0.31 \leq z/R_{AIP} \leq 0.95$). Controlled cases correspond to OC-3 for $Ma_{AIP}=0.36,0.46$ and OC-11 for $Ma_{AIP}=0.54$. In (b), solid and dashed lines denote uncontrolled and controlled cases, respectively; only blue curves are labeled.}\label{Fig20}
\end{figure}

 \subsection{Effect of control actuation on Reynolds stress events }

Previously, in Fig. \ref{Fig19}, the effect of flow control on the variance of the pressure fluctuations and turbulent kinetic energy was investigated. Similarly, here we examine the effect of control on the Reynolds-stress events, particularly the ejection and sweep events that were shown to correlate with the local minima and maxima of total pressure, respectively.  Fig. \ref{Fig20}(a,c,e) compares the time histories of $u^{\prime}w^{\prime}$ for the controlled and uncontrolled cases at $(y/R_{AIP},z/R_{AIP})=(0,0.71)$ on the AIP for the three Mach numbers. For $Ma_{AIP}=0.36$ and $0.46$, opposition control reduces the magnitude of $u^{\prime}w^{\prime}$ by orders of magnitude, and fluctuations become nearly negligible. For $Ma_{AIP}=0.54$, the magnitude is also substantially reduced, although a few intense events remain, particularly around $t\in [0.162,0.163]$. 

The magnitudes of the ejection and sweep events for the controlled and uncontrolled cases along Rake 1 at the AIP, which extends vertically from the center toward the top wall, i.e., $y/R_{AIP}=0$ and $0.3 \leq z/R_{AIP} \leq 0.95$, are shown in Fig. \ref{Fig20}(b,d,f). The controlled and uncontrolled cases are denoted by dashed and solid lines, respectively. Across all three Mach numbers, the magnitudes of both the ejection and sweep events decrease by approximately an order of magnitude over most of Rake 1. This indicates that steady opposition control substantially attenuates the Reynolds-stress events associated with the extrema of the total pressure. This along with Fig. \ref{Fig19} shows that the steady opposition control is also able to reduce the fluctuations and more specifically the extreme events associated with distortion.

\bibliography{jfmbib}

\end{document}